\documentstyle[12pt]{article}
\begin{document}
\begin{titlepage}
\vspace{-5.0 cm}
\title{
\begin{flushright}
{\normalsize UUITP 12/1994   \\
\vspace{-0.3 cm}
hep-th/9406xxx }
\end{flushright}
\bigskip   \bigskip
{\bf An involution and dynamics for the $q$-deformed quantum top}}
\author{A.Yu.Alekseev
\thanks{Supported by Swedish Natural Science Research Council
(NFR) under the contract F-FU 06821-304;
on leave of absence from Steklov Mathematical Institute,
Fontanka 27, St.Petersburg, Russia}
\\
Institute of Theoretical Physics, Uppsala University,
\\ Box 803 S-75108, Uppsala, Sweden.\\ \\
L.D.Faddeev
\\
Steklov Mathematical Institute, 191011 \\
  Fontanka 27, St.Petersburg, Russia
\\
 and
\\
Research Institute for Theoretical Physics \\
Siltavourenpengar 20 C,   Helsinki, Finland
}
\date{June 1994}
\maketitle
\thispagestyle{empty}
\vspace{-0.7 cm}
\begin{abstract}
This preprint is the English translation of  our paper published in
Russian
 in the
issue of  St.-Petersburg journal Zapiski LOMI dedicated to the
jubilee
of Prof. O.A. Ladyzhenskaya in the beginning of 1993. We hope that
this text still may be of interest for specialists.

It is known that the involution corresponding to the compact form is
incompatible with comultiplication for quantum groups at $|q|=1$. In
this paper we consider the quantum algebra of functions on the
deformed space $T^{*}G_{q}$ which includes both the quantum group and
the quantum universal enveloping algebra as subalgebras. For this
extended object we construct an anti-involution which reduces to the
compact form  $*$-operation in the limit $q\rightarrow 1$. The
algebra of functions on  $T^{*}G_{q}$ endowed with the $*$-operation
may be viewed as an algebra of observables of a quantum mechanical
system.  The most natural interpretation for such a system is   a
deformation of the quantum symmetric top. We suggest a discrete
dynamics for this system which imitates the free motion of the top.
\end{abstract}
\end{titlepage}

\section*{Introduction}
One year ago we have introduced the quantum dynamical system
$T^{*}G_{q}$
which is a deformation of  the  quantum top \cite{1}. In the
nondeformed case
the phase space of the symmetric top is a cotangent bundle of the
group of rotations $T^{*}G$ (the simplest case being $G=SU(2)$).  The
same phase space
describes a point particle moving in the group $G$. After
quantization the
space of states ${\cal H}$ of the dynamical system under
consideration can
 be naturally realized in square
integrable functions on the group (in the space $L^{2}(G)$).  The
regular representation of the
group $G$ acts in the space $L^{2}(G)$. So, the top is a natural
dynamical
system closely related to representation theory.

After  deformation the group $G$ is replaced by the corresponding
quantum group \cite{2}. Parameter of the deformation plays an
important role in the theory ( for example, dimension of  the space
of states essentially depends of  its value) and is denoted
traditionally by $q$. Introduction of $T^{*}G_{q}$ was stimulated by
the problem of  construction of  differential calculus on  quantum
groups \cite{3}-\cite{6}.  Unexpectedly, the same system  appeared in
the Wess-Zumino-Novikov-Witten model in  conformal field  theory
\cite {7}-\cite{9}. As a result the role of quantum groups in
conformal field theory was significantly clarified.

All constructions in \cite{1} are performed in the case of complex
coordinates on the group and for a complex deformation parameter. For
applications  the case where the group is compact and the value of
$q$ lies on the unit circle of the complex plane  is  the most
interesting one.   Thus, for  $ |q| =1$ the  problem of constructing
an anti-involution in the algebra of observables on  $T^{*}G_{q}$
arises.  An anti-involution  singles out  some real form of the
algebra of observables. In the classical limit
$q\rightarrow 1$ this real form must coincide with the algebra of
observables on $T^{*}K$, where $K$ is a compact form of  the group
$G$. Such an anti-involution will be constructed in this paper.
Moreover, we make use of the case and present a dynamical system on
$T^{*}G_{q}$ which is the most precise quantum group analog of the
symmetric top.

Following a common practice we shall write formulas, which can be
used for any group $G$, but we shall illustrate them by the example
of the group $SL(2)$.

In Section 1   we  give a background concerning $T^{*}G_{q}$.
The exposition follows \cite{1}.
In Section 2 we  introduce and discuss an  anti-involution which
corresponds to  the compact form of the group in the classical limit.
In Section 3 we describe  dynamics of the symmetric top on a quantum
group.

\section{Algebra of observables on $T^{*}G_{q}$}

Let $G$ be a classical simple Lie group, $q$ a complex parameter,
$R_{+}(q)$ a corresponding $R$-matrix \cite{10} in the fundamental
representation $V$. In the case  of  $G=SL(2)$ the fundamental
representation is two-dimensional, $V=C^{2}$,  and $R_{+}(q)$ is a
$4\times 4$
matrix. In the natural basis for $C^{2}\times C^{2}$ the $R$-matrix
is of the form\begin{equation}
R_{+}(q)=
\left (\matrix{
q^{1/2} & 0 & 0 & 0 \cr
0 & q^{-1/2}& q^{1/2}-q^{-3/2} & 0 \cr
0 & 0 & q^{-1/2} & 0 \cr
0 & 0 & 0 & q^{1/2}\cr
}\right ) \ .
\end{equation}

We omit the argument  $q$ in the notation of $R$-matrix.  Further we
assume that all $R$-matrices are calculated for  the same value of
$q$, which enters into the definition of  the theory $T^{*}G_{q}$
as a parameter.

Along with $R_{+}$ it  is useful to introduce one more $R-$matrix
\begin{equation}
R_{-}=P(R_{+})^{-1}P,
\end{equation}
where $P$ is a permutation in the tensor product $V\otimes
V(Pa\otimes b=b\otimes a).$

Coordinates on $T^{*}G_{q}$ or,  more precisely, generators of the
algebra of functions on the noncommutative manifold $T^{*}G_{q}$ may
be combined into matrices $g$ and $\Omega_{\pm}$. The matrix $g$ is a
quantum analog of a group element. Its matrix elements are
coordinates on the  base of the quantum bundle $T^{*}G_{q}$. The
matrices $\Omega_{+}$ and $\Omega_{-}$ are chosen as upper-  and
lower-triangular, respectively. Moreover, their diagonal parts
$\omega_{+}$ and $\omega_{-}$ are inverse to each other
 $\omega_{+}\omega_{-}=\omega_{-}\omega_{+}=1$. The entries of
$\Omega_{+}$ and $\Omega_{-}$ are coordinates in a fiber of  the
quantum bundle $T^{*}G_{q}$.

In the simplest case of $G=SL(2)$ one can write $g$ and
$\Omega_{\pm}$ in components
\begin{center}
$g=\pmatrix{
a & b \cr
c & d\cr
} , \;\;\; det_{q}g=ad-qbc=1; $
\end{center}
\begin{equation}
\Omega_{+}=
\pmatrix{
K^{1/2} & (q^{3/2}-q^{-1/2})X_{+} \cr
0 & K^{-1/2} \cr
},\, \Omega_{-}=\pmatrix{
K^{-1/2} & 0 \cr
-(q^{1/2}-q^{-3/2})X_{-} & K^{1/2} \cr
} \ .
\end{equation}

The matrix elements of $\Omega_{\pm}$ are expressed through the
generators $K,X_{+},X_{-}$ of the quantum algebra $sl_{q}(2) $
\cite{10}.  For more detail, we refer the reader to papers
\cite{1},\cite{10}.  Here we write the commutation relations for the
entries of $g,\Omega_{\pm}$. They form  a quadratic algebra, it is
convenient to write this algebra in the form \cite{1}:
\begin{equation}
Rg^{1}g^{2}=g^{2}g^{1}R,
\end{equation}

\begin{eqnarray}
R\Omega_{+}^{1}\Omega_{+}^{2}=\Omega_{+}^{2}\Omega_{+}^{1}R,
\nonumber \\
R\Omega_{-}^{1}\Omega_{-}^{2}=\Omega_{-}^{2}\Omega_{-}^{1}R, \\
R_{+}\Omega_{+}^{1}\Omega_{-}^{2}=\Omega_{-}^{2}\Omega_{+}^{1}R_{+},
\nonumber
\end{eqnarray}

\begin{eqnarray}
R_{+}\Omega_{+}^{1}g^{2}=g^{2}\Omega_{+}^{1}, \nonumber \\
R_{-}\Omega_{-}^{1}g^{2}=g^{2}\Omega_{-}^{1}.
\end{eqnarray}
Here we use the notations from \cite{10}, which allow us to write
the commutation relations of all entries of two matrices in a single
formula. Namely, for any matrix $A$ acting in the space $V$ one can
construct the matrices

\begin{center}
$A^{1}=A\otimes I,\;\; A^{2}=I\otimes A$
\end{center}
in the space $V\otimes V$.  Then one can understand formulas (4-6) as
relations for matrices acting in $V\otimes V$ taking into account the
operator order of factors (for example, in the left-hand side of (6)
the entries of  $\Omega_{\pm}$ are situated to the left of the
entries of $g$, and in the right-hand side to the right).

In the first three relations one can use either $R_{+}$ or $R_{-}$ as
the matrix $R$. In the fourth we must use $R_{+}$.  As an alternative
we write the relation
\begin{center}
$R_{-}\Omega_{-}^{1}\Omega_{+}^{2}=
\Omega_{+}^{2}\Omega_{-}^{1}R_{-},$
\end{center}
which is obtained upon multiplication of (5)  by $P$ from the left
and from the right.
Here we use the relation
\begin{center}
$PA^{1}P=A^{2}$
\end{center}
and  equality (2).

It is easy to check  that the form of matrices $g,\Omega_{\pm}$ in
(3) (the matrices $\Omega_{\pm}$ are triangular and the
$q$-determinant of $g$ is equal to 1) is compatible with relations
(4-6).

The subalgebras of the algebra of functions on $T^{*}G_{q}$ generated
by the entries of  $g$ and $\Omega_{\pm}$ respectively, are Hopf
algebras. The algebra generated by the entries of $g$ is known as the
algebra of functions on the quantum group $Funk_{q}(G)$, and the
algebra generated by the entries of $\Omega_{\pm}$ as the quantized
universal enveloping algebra $U_{q}({\cal G})$ where ${\cal G}$ is
the Lie algebra corresponding to the group $G$. Coproduct in these
Hopf algebras is given by simple formulas \cite{10}:
\begin{eqnarray}
\Delta g^{ik}=\sum_{j}  (g^{'})^{ij} (g^{''})^{jk}, \nonumber \\
\Delta\Omega_{\pm}^{ik}=\sum_{j}
(\Omega_{\pm}^{'})^{ij}(\Omega_{\pm}^{''})^{jk}.
\end{eqnarray}
The full  algebra of functions on $T^{*}G_{q}$ does not possess  the
structure of a Hopf algebra.

As is first observed in \cite{11}, it is convenient to use the matrix
\begin{equation}
\Omega=\Omega_{+}\Omega_{-}^{-1},
\end{equation}
where $\Omega_{-}^{-1}$ is the antipode of the matrix $\Omega_{-}$.
In the case of  $G=SL(2)$, the matrix $\Omega_{-}^{-1}$ has the
following form:

\begin{center}
$\Omega_{-}^{-1}=\pmatrix{
K^{1/2}& 0 \cr
(q^{3/2}-q^{-1/2})X_{-} & K^{-1/2} \cr
}.$
\end{center}
The commutation relations for $\Omega_{\pm}$ and $g$ can be rewritten
for $\Omega$ in the form
\begin{equation}
\Omega^{1}(R_{-})^{-1}\Omega^{2}R_{-}=
R_{+}^{-1}\Omega^{2}R_{+}\Omega^{1}
\end{equation}
and
\begin{equation}
R_{-}g^{1}\Omega^{2}=\Omega^{2}R_{+}g^{1}.
\end{equation}

Let us consider the classical limit of the relations (9-10). We set
\begin{equation}
q=e^{i\hbar\gamma},
\end{equation}
where $\gamma$ is a parameter of deformation and $\hbar$ is the
Planck constant  which controls the passage  from classical  to
quantum mechanics.
One commonly uses $\hbar=1$, but, by methodical considerations, it is
convenient to preserve $\hbar$ as an independent constant in the
theory. For small $\gamma$, we can expand the $R$-matrices $R_{+}$
and $R_{-}$ in the series with respect to $\gamma$:
\begin{equation}
R_{\pm}=I + i\hbar \gamma r_{\pm}+\ldots \ ,
\end{equation}
where $r_{\pm}$ are classical  $r-$matrices. For example, in the case
$G=SL(2)$
\begin{eqnarray}
r_{+}=\left (\matrix{
1/2 & 0 & 0 & 0 \cr
0 & -1/2 & 2 & 0 \cr
0 & 0 & -1/2 & 0 \cr
0 & 0 & 0 & 1/2 \cr
}\right ), \nonumber \\
r_{-}=-Pr_{+}P=\left (\matrix{
-1/2 & 0 & 0 & 0 \cr
0 & 1/2 & 0 & 0 \cr
0 & -2 & 1/2 & 0 \cr
0 & 0 & 0 & -1/2 \cr
}\right ).
\end{eqnarray}
The difference  of the classical $r$-matrices
\begin{equation}
C=r_{+}-r_{-}
\end{equation}
coincides with the tensor Casimir operator for Lie algebra ${\cal
G}$. For example, for  ${\cal G}=sl(2)$
\begin{equation}
C=\left (\matrix{
1 & 0 & 0 & 0 \cr
0 & -1 & 2 & 0 \cr
0 & 2 & -1 & 0 \cr
0 & 0 & 0 & 1 \cr
}\right )=
\sum_{a=1}^{3}{\sigma^{a}\otimes\sigma^{a}},
\end{equation}
where $\sigma^{a}$ are Pauli matrices.

In the classical limit $\gamma\rightarrow0$ we set
\begin{equation}
\Omega=I+\gamma\omega+\ldots \  .
\end{equation}
Then, by setting $\gamma=0$, we have the following relations for the
matrices $g$ and $\omega$:
\begin{eqnarray}
g^{1}g^{2}=g^{2}g^{1} \ , \nonumber \\
\omega^{1}\omega^{2}-\omega^{2}\omega^{1}=-i\hbar [ C ,\omega^{2}] \
, \\
g^{1}\omega^{2}-\omega^{2}g^{1}=i\hbar C g^{1} \ . \nonumber
\end{eqnarray}
By setting $\hbar\rightarrow 0$, we obtain the Poisson brackets for
the entries of $g$ and $\omega$:
\begin{eqnarray}
\{ g^{1},g^{2} \} =0, \nonumber \\
\{ \omega^{1}, \omega^{2} \} =\frac{1}{2}[ C,\omega^{1}-\omega^{2} ],
 \\ \nonumber
    [  \omega^{1},g^{2}  ] = -Cg^{2}.
\end{eqnarray}
It is easy to check that the Poisson brackets (18) coincide  with the
canonical Poisson structure on  the manifold $T^{*}G$, where $g$ is a
group element and $\omega$ is a right-invariant  momentum chosen as a
coordinate on the fiber.

This completes the short review of main properties of the model
$T^{*}G_{q}$ and we pass to the informative part of the paper.

\section{An anti-involution for $T^{*}G_{q}$}

We assume that $q$ satisfies the relation
\begin{equation}
qq^{*}=1.
\end{equation}
In that case the $R-$matrices $R_{+}$ and $R_{-}$ are Hermitian
conjugate with each other
\begin{equation}
(R_{+})^{*}=R_{-}.
\end{equation}
It is easy to introduce  an operation * for the matrices
$\Omega_{\pm}$ compatible with relations (5).  Namely, we can put
\begin{equation}
\Omega_{+}^{*}=\Omega_{-},\;\;\;\Omega_{-}^{*}=\Omega_{+}.
\end{equation}
In terms   of generators of the quantum algebra we have a $q-$analog
of the $SU(2)$-involution:
\begin{equation}
K^{*}=K^{-1},\;\;\; X_{+}^{*}=X_{-},\;\;\;X_{-}^{*}=X_{+}.
\end{equation}
In the classical limit (16)  formulas (21) turn into the  relation
\begin{equation}
\omega^{*}_{+}=\omega_{-},\;\;\; \omega_{-}^{*}=\omega_{+}.
\end{equation}
For the complete matrix

\begin{center}
$\omega=\omega_{+}-\omega_{-}$
\end{center}
we see that $\omega$ is anti-Hermitian:
\begin{equation}
\omega^{*}=-\omega.
\end{equation}
Condition (24) singles out the compact form of  the algebra ${\cal
G}$. It becomes clear that the similar condition for the group
\begin{equation}
g^{*}=g^{-1}
\end{equation}
is not compatible with  relation (4) and it can not be used in the
definition of an anti-involution for $T^{*}G_{q}$. The rest of the
section is devoted to the description of a way out  the situation and
to the construction of  $g^{*}$.

In order to describe an alternative to (25) we need several new
objects. It is convenient to consider, together with the
right-invariant momentum $\Omega_{L}$ generating left  translations
on the group, the left-invariant momentum $\Omega_{R}$ generating
right translations.
$\Omega_{R}$ is expressed through $\Omega_{L}$  by means of the
simple formula
\begin{equation}
\Omega_{R}=g^{-1}\Omega_{L}g.
\end{equation}
It is easy to check that $\Omega_{R}$ and $\Omega_{L}$ commute,
\begin{equation}
\Omega_{L}^{1}\Omega_{R}^{2}=\Omega_{R}^{2}\Omega_{L}^{1}.
\end{equation}
By analogy with $\Omega_{L}$, we decompose  $\Omega_{R}$  into a
product of upper- and lower-triangular matrices with equal diagonal
parts (we mean $\Sigma_{+}$ and $\Sigma_{-}^{-1}$)
\begin{equation}
\Omega_{R}=\Sigma_{+}\Sigma_{-}^{-1},
\end{equation}
so that
\begin{equation}
g^{-1}\Omega_{\pm}=\Sigma_{\pm}h.
\end{equation}
The coordinates $\Sigma_{\pm}$ and $h$ yield an alternative
parametrization on $T^{*}G_{q}$. In  the classical limit
$\gamma\rightarrow 0$ the matrices $\Omega_{\pm}$ and $\Sigma_{\pm}$
tend to $1$, and we get
\begin{equation}
g^{-1}=h.
\end{equation}
The required  anti-involution is of the form
\begin{equation}
g^{*}=h.
\end{equation}
It turns out to be compatible with  basic relations (4-6). Observe
that  the components of  the right momentum $\Omega_{R}$
\begin{equation}
\Sigma_{\pm}=g^{-1}\Omega_{\pm}h^{-1}
\end{equation}
behave in the same way  as the components of  the left momentum:
\begin{equation}
\Sigma_{+}^{*}=\Sigma_{-},\;\;\;\Sigma_{-}^{*}=\Sigma_{+}
\end{equation}
(cf. formula (21)).

In the algebra of functions on $T^{*}G_{q}$ relations (21) and (30)
determine an anti-involution compatible with relations (4-6). This
allows us to say that we have constructed a $q$-analog of the
cotangent bundle of a compact group (in the simplest case, of the
group $SU(2)$) for $|q|=1$. We observe that  in this framework an
analog of the compact group itself is not well defined, since the
involution (25) is not compatible with relations (4). More precisely,
as is easy to see, $g^{*}$ and $g^{-1}$ have different commutation
relations and we can not make them equal without contradictions.
Thus, the main conclusion is that, despite of lack of $q$-analogs of
compact groups for $|q|=1$, one can successfully construct
$q$-analogs of their cotangent bundles, which is to a large extent an
adequate alternative.

In conclusion of this section, we introduce a set of relations for
$\Sigma_{\pm}$ and $h$:
\begin{center}
$
R_{\pm}h^{1}h^{2}=h^{2}h^{1}R_{\pm}; $
\vspace{3 mm}

$\Sigma_{+}^{1}\Sigma_{+}^{2}R_{\pm}=
R_{\pm}\Sigma_{+}^{2}\Sigma_{+}^{1}, $

$\Sigma_{-}^{1}\Sigma_{-}^{2}R_{\pm}=
R_{\pm}\Sigma_{-}^{2}\Sigma_{-}^{1}, $

$\Sigma_{+}^{1}\Sigma_{-}^{2}R_{+}=
R_{+}\Sigma_{-}^{2}\Sigma_{+}^{1}; $
\vspace{3 mm}

$h^{1}\Sigma_{+}^{2}=\Sigma_{+}^{2}R_{-}h^{1}, $

$h^{1}\Sigma_{-}^{2}=\Sigma_{-}^{2}R_{+}h^{1},$
\vspace{3 mm}

$\Omega^{1}\Sigma^{2}=\Sigma^{2}\Omega^{1}; $
\vspace{3 mm}

$g^{1}R_{-}\Sigma_{+}^{2}=\Sigma_{+}^{2}g^{1}, $

$g^{1}R_{+}\Sigma_{-}^{2}=\Sigma_{-}^{2}g^{1}; $

\vspace{3 mm}
$h^{1}\Omega_{+}^{2}=\Omega_{+}^{2}h^{1}R_{-},$

$h^{1}\Omega_{-}^{2}=\Omega_{-}^{2}h^{1}R_{+}. $

\end{center}

We left to the reader interested in this topic to verify the
consistency of this system of relations and its compatibility with
the basic relations (4-6).

\section{Dynamics on $T^{*}G_{q}$.}

At the beginning we recall the structure of dynamics in the
nondeformed case. The classical symmetric top  is described by the
Lagrangian
\begin{equation}
L=Tr(\omega_{L}\dot{g}g^{-1}-\frac{1}{2}\omega^{2}_{L})=
Tr(\omega_{R}g^{-1}\dot{g}-\frac{1}{2}
\omega_{L}^{2}),
\end{equation}
where $g$ is a group element ,  $\omega_{L}$ and $\omega_{R}$ are
left and right momenta, respectively,  and
\begin{equation}
\omega_{R}=g^{-1}\omega_{L}g.
\end{equation}
The equations of motion are of the form
\begin{equation}
\dot{g}=\omega_{L}g, \;\;\;\dot{\omega}_{L}=0
\end{equation}
or
\setcounter{equation}{35}
\begin{equation}
\dot{g}=g\omega_{R},\;\;\;  \dot{\omega}_{R}=0.
\end{equation}
The solution of the equations (36) can written as follows
\begin{eqnarray}
\omega_{L}(t)=\omega_{L}(0)=\omega_{L}, \nonumber \\
g(t)=e^{\omega_{L}t}g(0)
\end{eqnarray}
and it gives rise to the mapping $(g(0),\omega(0))\rightarrow
(g(t),\omega(t)),$  preserving the Poisson structure on $T^{*}G$.

We would like to construct  an analog of   such a mapping for the
deformed case.
Such an analog  does exist but one can not  make the time variable
$t$ continuous in a natural way. It turns out to be discrete and
takes say integer values $t=0,1, 2,\dots$ and our dynamical system
turns into a cascade. The values of the variables $g(t)$ and $\Omega
(t)$ are given by the formulas
\begin{eqnarray}
\Omega(n)=\Omega(0), \nonumber \\
g(n)=\Omega^{n}g(0).
\end{eqnarray}
The main property of the transformations (38) consists in preserving
(4), (9), (10). It is sufficient to check that this property holds
for $n=1$. We make these simple calculations

\vspace{5mm}
\hspace{25mm} $\Omega^{1}g^{1}\Omega^{2}g^{2}=$ \hspace{27 mm} (by
formula (10))

\hspace{25 mm} $=\Omega^{1}(R_{-})^{-1}\Omega^{2}R_{+}g^{1}g^{2}=$
\hspace {5 mm} (by formulas (4) and (9))

\hspace{25 mm}$=R_{+}^{-1}\Omega^{2}
R_{+}\Omega^{1}R_{-}^{-1}g^{2}g^{1}R_{+}=$ \hspace{1 mm} (using (10))

\hspace{25 mm}
$=R_{+}^{-1}\Omega^{2}R_{+}R_{+}^{-1}g^{2}\Omega^{1}g^{1}R_{+}=$

\hspace{25mm} $=R_{+}^{-1}\Omega^{2}g^{2}\Omega^{1}g^{1}R_{+}.$

\vspace{5mm}
which coincides with relation (4):
\begin{center}
$R_{+}\Omega^{1}g^{1}\Omega^{2}g^{2}=
\Omega^{2}g^{2}\Omega^{1}g^{1}R_{+}.$
\end{center}

Further we  consider the expression

\vspace{5mm}
\hspace{30 mm}$R_{-}\Omega^{1}g^{1}\Omega^{2}= $\hspace{17 mm} (using
(10))

\hspace{30 mm}$=R_{-}\Omega^{1}R_{-}^{-1}\Omega^{2}R_{+}g^{1}=$
(using (9))

\hspace{30 mm}$=\Omega^{2}R_{+}\Omega^{1}R_{+}^{-1}R_{+}g^{1}=$

\hspace{30 mm}$=\Omega^{2}R_{+}\Omega^{1}g^{1}$,

\vspace{5mm}
\noindent which reproduces the formula (10).

Thus, we have verified the basic relations for $n=1$. By induction,
the same is true for any natural $n$.

Observe one more important property of the evolution (38). Namely, it
is compatible with the anti-involution proposed in the previous
section. This means that if we start with $\Omega_{\pm}(0)$ and
$g(0)$ such that
\begin{center}
$\Omega^{*}_{+}(0)=\Omega_{-}(0),\;\;\; g^{*}(0)=h(0),$
\end{center}
then  $\Omega_{\pm}$ and $g$ will satisfy these relations at any time
$t=n$.  We check this assertion. First, we have
\begin{center}
$\Omega_{\pm}(n)=\Omega_{\pm}(0)$
\end{center}
and so the relations for $\Omega_{\pm}(n)$ are valid automatically.

We again  check the relations for $g$ only for $n=1$. In that case
\begin{eqnarray}
g^{-1}(1)\Omega_{+}=\Sigma_{+}(1)h(1), \nonumber \\
g^{-1}(1)\Omega_{-}=\Sigma_{-}(1)h(1).
\end{eqnarray}
Using  relations (8), (29) and (38), it is easy to evaluate
$\Sigma_{\pm}(1)$ and $h(1)$. They are equal to
\begin{equation}
\Sigma_{\pm}(1)=\Sigma_{\pm}(0), \;\;\;
h(1)=h(0)\Omega_{+}^{-1}\Omega_{-}.
\end{equation}
Now  it remains to find the relation between  $g(1) $ and $h(1)$.
Substituting (38) to (40) we obtain
\begin{eqnarray}
g^{*}(1)=g^{*}(0)\Omega^{*}=h(0)(\Omega_{+}\Omega_{-}^{-1})^{*}=
\nonumber \\
=h(0)(\Omega_{-}^{*})^{-1}\Omega_{+}^{*}=
h(0)\Omega_{+}^{-1}\Omega_{-}=h(1).
\end{eqnarray}
Thus, the discrete evolution (38) is a natural analog of the
continuous evolution (37). It preserves the commutation relations of
the algebra $T^{*}G_{q}$ and is compatible with the anti-involution
that singles out the compact form.

This completes the main part of the paper and we pass to discussion
of applications and perspectives.

\section*{Conclusions}

This paper may have meaning for several  directions connected with
the quantum group theory. We discuss two opportunities which, to our
opinion, are of most interest.

{\bf 1}. As was shown in Section 3, one can naturally introduce a
discrete evolution on  $T^{*}G_{q}$ in such a way that $t$ takes
integer values. At the same time, to introduce $g(t)$ for arbitrary
$t$, or, equally, to introduce fractional powers of $\Omega$, does
not seem to be natural. In other words, deformation is clearly
related to the discreteness of parameters of one-parameter
automorphism groups.

Omitting here the discussion of possible physical applications, we
rewrite formula (38) for $t=1$ in the form

\begin{equation}
\Omega=g(1)g(0)^{-1}.
\end{equation}
This formula is a $q-$analog of the classical formula
\begin{center}
$\omega=dg\;g^{-1}$
\end{center}
for the Maurer-Cartan form on the group $G$. Here we  approach the
problem of constructing of  differential calculus on a quantum group.
This problem is discussed by many authors \cite{3}-\cite{6}.  It
seems to us that the fact that the differential (42) is not naively
infinitesimal is not understood as yet.  So, in the paper \cite{6}
B.Zumino makes use of the object
\begin{center}
$X=\Omega-I$,
\end{center}
which in the classical limit $\gamma\rightarrow 0$ after
renormalization tends to $\omega$ but it  might be not the most
natural way in the deformed case. As one of the authors of (LDF) (for
example, at the seminar "Quantum Groups" of  the  Euler International
Mathematical  Institute, in the Fall 1990) repeatedly stated, he is
skeptical of the  construction of a differential on a quantum group
with naive Leibniz rule.

We hope to return to  application of the technique that we have
developed here to differential calculus on quantum groups.

{\bf 2}. The system $T^{*}G_{q}$ with $q$ being a root of unity,
$q^{N}=1$, is most interesting for physical applications. For
example, $q=e^{2\pi i/N}$. In that case the algebra $U_{q}({\cal G})$
has only finitely many irreducible representations for which the
theory of  tensor products is closely related to such an area of
modern mathematical physics as rational conformal field theory
(RCFT). Following the analogy with the nondeformed case one can
assume that the algebra $T^{*}G_{q}$ can be represented in the
$q$-analog of the regular representation  of $U_{q}({\cal G})$. We
may assume with good reason that for $q$ being a root of unity  the
regular representation turns out to be finite dimensional. Thus, an
interesting problem consists in explicit constructing of the finite
dimensional $*$-representation of the algebra of functions on
$T^{*}G_{q}$. The properties of such a representation will model the
main features of RCFT on a finite dimensional example and essentially
explain them  \cite{7,8}.

\subsection*{Note Added}

Since the Russian version of this paper had been written, there is
some progress in the field and we acknowledge here some results and
add some references missed in the text.

{\bf 1}.  An anti-involution of the same type as the one constructed
in this paper has been earlier independently considered in \cite{12}.
This concept has been applied for general quadratic $R$-matrix
algebras parametrized by graphs in \cite{13}.

{\bf 2}. Differential calculus which satisfy a modified Leibniz rule
and appropriate for the $A$-series $G=SL(n)$ has been suggested in
\cite{14}.

{\bf 3}. The regular representation of $T^{*}G_{q}$ will be
considered among other topics in the paper \cite{15}.

\end{document}